\documentclass[pra,twocolumn,showpacs,preprintnumbers,superscriptaddress]{revtex4}%

\usepackage{graphicx}
\usepackage{dcolumn}
\usepackage{bm,hyperref}
\usepackage{amsmath}
\usepackage{amsfonts}
\usepackage{amssymb}%
\newlength{\myx} 
\newlength{\myy} 
\newcommand\includegraphicstotab[2][\relax]{%
\settowidth{\myx}{\includegraphics[{#1}]{#2}}%
\settoheight{\myy}{\includegraphics[{#1}]{#2}}%
\parbox[c][1.1\myy][c]{\myx}{%
\includegraphics[{#1}]{#2}}%
}

\providecommand{\U}[1]{\protect\rule{.1in}{.1in}}
\begin{document}

\title{An advanced Jones calculus for the classification of periodic metamaterials}
\author{Christoph Menzel}
\affiliation{Institute of Condensed Matter Theory and Solid State
Optics, Friedrich-Schiller-Universit\"at Jena, Max-Wien-Platz 1,
D-07743 Jena, Germany}

\author{Carsten Rockstuhl}
\affiliation{Institute of Condensed Matter Theory and Solid State
Optics, Friedrich-Schiller-Universit\"at Jena, Max-Wien-Platz 1,
D-07743 Jena, Germany}

\author{Falk Lederer}
\affiliation{Institute of Condensed Matter Theory and Solid State
Optics, Friedrich-Schiller-Universit\"at Jena, Max-Wien-Platz 1,
D-07743 Jena, Germany}

\begin{abstract}
By relying on an advanced Jones calculus we analyze the polarization properties
of light upon propagation through metamaterial slabs in a comprehensive
manner. Based on symmetry considerations, we show that all periodic
metamaterials may be divided into five different classes only. It is shown that
each class differently affects the polarization of the transmitted light and
sustains different eigenmodes. We show how to deduce these five classes from
symmetry considerations and provide a simple algorithm that can be applied to
decide by measuring transmitted intensities to which class a given metamaterial
is belonging to only. \end{abstract}

\pacs{XX.XX.XX} \maketitle

\section{Introduction}
Metamaterials (MM) provide a large variety of unprecedented optical properties.
Whereas in its infancy properties such as a dispersive permeability were at the
focus of interest \cite{ScienceLinden,EnkrichPRL,ScienceSoukoulis}, the range of
properties to be intentionally affected by suitably chosen MMs significantly
increased. More and more complex
\cite{PadillaPRB,WallpaperPadilla,DeckerCouplingSRR,Stereometamaterials} and
most notably, chiral
\cite{Rogacheva2006,ShuangPRL,PlumNegChiral2009,ZhouNegChiral,ReviewWang,ChiralRockstuhl,Rotator2010}
or quasi-planar \cite{Bai2007,APLChiral,Jefimovs2008} chiral structures
attracted a great deal of attention due to their polarization selective optical
response and their potential to implement functional devices with
unprecedented applications such as, e.g., broadband polarizers for circular light
\cite{ScienceGoldHelix}. Moreover, from a scientific point of view, such complex
MMs permit to observe unexpected and counterintuitive effects like asymmetric
transmission for circular
\cite{FedotovPRL,NanoLettFedotov,NanoLettSchwanecke,PRBRanjan} or even
for linearly polarized light \cite{PRLMenzel}.

Recent studies have shown that the assignment of effective material parameters
is doubtful in many cases \cite{Validity2010} and generally requires the
assumption of complex constitutive relations
\cite{PadillaOE,Arnaut1997,SerdyukovBook}. Thus a more suitable target
function to be tailored by an appropriate MM design is the optical response
itself. This optical response is completely involved in the response functions,
such  as complex reflection and transmission coefficients, for a given input
illumination. This paradigm change reflects that for an actual application a
certain value of some effective material coefficient is of minor importance, as
long as the sample exhibits the desired optical response.

The response, in particular in transmission, can then be easily described by transmittances and
polarization ellipses, averaged polarization rotation and polarization conversion \cite{Rogacheva2006,ChiralRockstuhl,PlumAPL}. These
phenomenological quantities can be completely determined from the frequency dependent Jones
matrix \cite{Jones1941} that relates the complex amplitudes of the incident to the transmitted field. We will
term this Jones matrix throughout the manuscript the T-matrix since it fully describes how
the light is transmitted through a metamaterial slab. This 2 $\times$ 2 matrix
comprises, in general, four different complex and dispersive quantities, reflecting the
spectral properties of the MM. The associated Jones calculus can be applied to describe the
transmission of an arbitrarily polarized incident plane wave through a MM slab if only the
zeroth order Bloch mode emerges. This holds for MMs composed of periodically arranged sub-wavelength
unit cells and we will assume this throughout the manuscript. For the sake of simplicity we also assume
that the structures are symmetrically embedded. We  assume that all materials are linear
and reciprocal, i.e., excluding Faraday media. No further restrictions on the symmetry of the
unit cells and the generally complex permittivity of the constituting materials are necessary.

With this work we intend  to introduce a classification of periodic
metamaterials based on their symmetry properties and to link them to their specific
T-matrix. We will show that all metamaterials can be divided into
only five distinct classes, each having an individual form of the
T-matrix and specific eigenstates. Each of these five classes leads to a very specific transmission
characteristics directly linked to the symmetry of the structure.
Therefore, this investigation provides a useful tool to analyze the optical
response of complex MMs and it may serve as a guide to identify designs for a desired
polarization response. Although for fabricated MM the geometry is usually known, the
application of combinatorial approaches to explore new MM geometries in the near future
requires such tool to classify the properties of MMs.

The paper is structured as follows. In Sec. II we present the necessary fundamentals
to handle the generally complex valued T-matrices and derive general expressions for the
eigenpolarizations. In Sec. III we derive the form of the T-matrix for the most relevant
symmetry classes. In Sec. IV  we provide examples of metamaterials for these symmetry classes
and discuss briefly their optical behavior. In Sec. V a comprehensive tabular overview
is given to summarize the results and we present a simple scheme to classify MM samples without having
a priori knowledge in terms of the presented formalism by measured transmittances only.

\section{Basic Theory}
\begin{figure}[h]
\begin{center}
\includegraphics[width=84mm,angle=0] {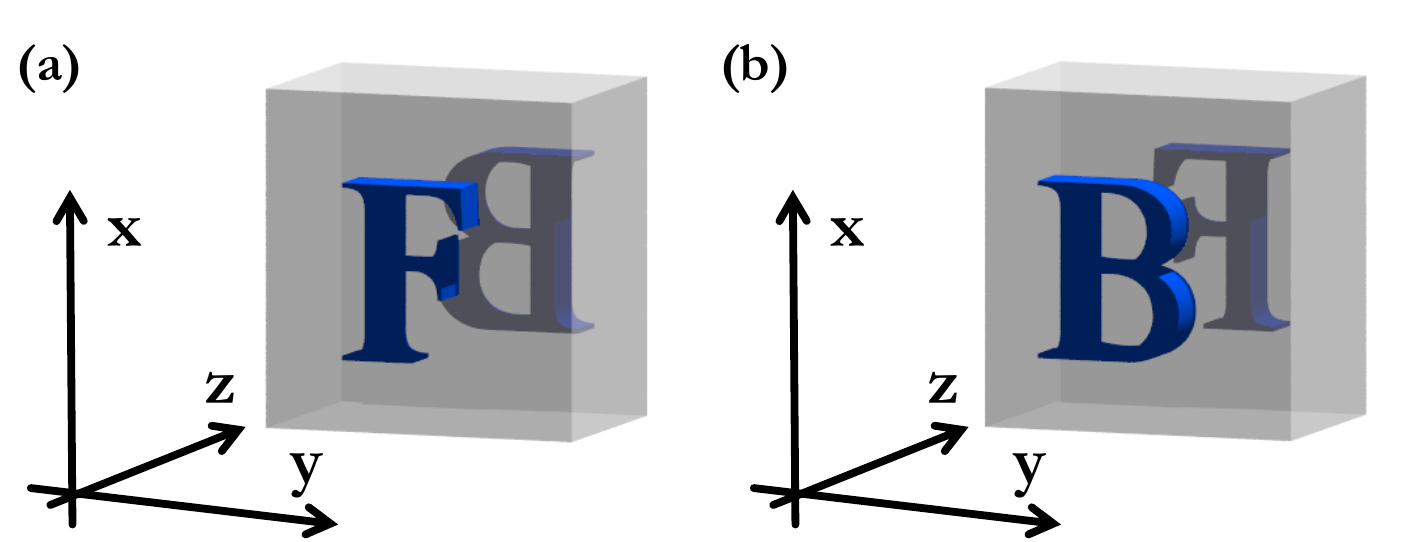}
\caption{Schematic of the geometry. a) and b) show the sample from
opposite sides with F and B indicating the front- and back-side , respectively.}
\label{FIG_KOS}
\end{center}
\end{figure}

It is assumed that the MM slab is illuminated by a plane wave propagating in positive $z$-direction
$$\mathbf{E}_i(\mathbf{r},t)=\begin{pmatrix}I_x \\ I_y\end{pmatrix}e^{i(kz-\omega t)},$$
with $\omega$ being its frequency, $k=\omega/c\sqrt{\epsilon(\omega)}$ the wavevector, and the complex amplitudes $I_x$ and $I_y$ describing the state of polarization.
The transmitted field is then given by
$$\mathbf{E}_t(\mathbf{r},t)=\begin{pmatrix}T_x \\ T_y\end{pmatrix}e^{i(kz-\omega t)},$$
where we have assumed that the medium is sandwiched between a medium characterized by the permittivity $\epsilon(\omega)$.
A sketch of the geometry is depicted in Fig.~\ref{FIG_KOS}. The unit cells are periodically
arranged in $x$- and $y$-direction without restricting to a particular lattice.
We assume coherent, monochromatic plane waves so to use a generalized
Jones calculus instead of the Mueller calculus necessary for incoherent light \cite{OEEbbesen,Chipman1996}.
The Jones calculus is said to be generalized since we allow for arbitrary
complex Jones matrices which we will call T-matrices (transmission matrices).

The T-matrix connects the generally complex amplitudes of the
incident and the transmitted
field:
\begin{equation}
\begin{pmatrix}T_x \\ T_y\end{pmatrix}=\begin{pmatrix}T_{xx} & T_{xy} \\ T_{yx} & T_{yy} \end{pmatrix}\begin{pmatrix}I_x
\\ I_y\end{pmatrix}=\begin{pmatrix}A & B \\ C & D \end{pmatrix}\begin{pmatrix}I_x \\ I_y\end{pmatrix}=\hat{T}^{\textrm{f}}\begin{pmatrix}I_x \\ I_y\end{pmatrix},\label{EQ_LinForward}
\end{equation}
where for convenience we have replaced the entries $T_{ij}$ by $A,B,C,D$  which form the
actual T-matrix. In the following few subsections we will discuss some generic properties
of this T-matrix.

\subsection{Directional dependent properties}
In the last term of Eq.~(\ref{EQ_LinForward}) the  T-matrix superscript f designates  propagation in
forward direction.  Of course, the choice of forward (f) and backward
(b) propagation is arbitrary. Thus $\hat{T}^{\textrm{b}}$ describes the
transmission matrix for light propagating through the structure
rotated by $180^\circ$ with
respect to the $x-$axis, where the choice of $x$ or $y$ is
arbitrary.

Since only reciprocal media are considered we have:
\begin{equation}
\hat{T}^b=\begin{pmatrix}A & -C \\ -B & D
\end{pmatrix},\label{EQ_LinBackward}
\end{equation}
where the minus sign in the off-diagonal elements accounts for
the rotation of the system looking from the backside \cite{Potton2004}. Therefore, the
complex matrix $\hat{T}^{\textrm{f}}$ already contains all information
necessary to determine light transmission for arbitrarily
polarized incident light from both main illumination directions.
Its is important to stress that this relation between $\hat{T}^{\textrm{f}}$ and $\hat{T}^{\textrm{b}}$
is in general only valid for this particular base where the coordinate
axis from the backside are given by replacing those of the frontside
by $x^{\textrm{b}}=\pm x^{\textrm{f}}$, $y^{\textrm{b}}=\mp y^{\textrm{f}}$. The actual sign depends on the
definition of the rotation of the system.

\subsection{Change of the base}

For analytical as well as for experimental concerns it is useful
to have at hand the transmission matrix in an arbitrary not necessarily orthogonal base. Let the vectors $\mathbf{i}$ and $\mathbf{t}$ denote the
incident and the transmitted light in a certain base. Then the
incident and the transmitted light in the Cartesian base is given
by $\mathbf{I}=\hat{\Lambda}\mathbf{i}$ and
$\mathbf{T}=\hat{\Lambda}\mathbf{t}$, respectively, with $\Lambda$ being the
change of basis matrix. Hence, the T-matrix for this new base is given by
\begin{equation}
\mathbf{T}=\hat{T}\mathbf{I}\,\rightarrow\,\mathbf{t}
=\hat{\Lambda}^{-1}\hat{T}\hat{\Lambda}\mathbf{i}=\hat{T}_{\mathrm{new}}\mathbf{i}={\begin{pmatrix}T_{11}
& T_{12} \\ T_{21} & T_{22} \end{pmatrix}}\begin{pmatrix}i_1 \\ i_2
\end{pmatrix}.\label{EQ_BaseTransform}
\end{equation}
All representations of the system are completely equivalent of
course. A transformation of practical importance is the change from the Cartesian base to the circular base. Then the change of basis matrix reads as
$$\hat{\Lambda}=\frac{1}{\sqrt{2}}\begin{pmatrix}1 & 1 \\ i & -i \end{pmatrix},$$
where the columns of the $\hat{\Lambda}$ matrix are the new
eigenstates. The T-matrix for circular states is then given by:
\begin{equation}
\hat{T}^{\mathrm{f}}_{\mathrm{circ}}=\begin{pmatrix}T_{++} & T_{+-} \\ T_{-+} &
T_{--}
\end{pmatrix}=\nonumber
\end{equation}
\begin{equation}
=\frac{1}{2}\begin{pmatrix}\left[A+D+i(B-C)\right] &
\left[A-D-i(B+C)\right] \\ \left[A-D+i(B+C)\right] &
\left[A+D-i(B-C)\right] \end{pmatrix},\label{EQ_T_CIRC_F}
\end{equation}
connecting the amplitudes of circularly polarized incident and transmitted light:
$$\begin{pmatrix}T_{+} \\ T_{-}\end{pmatrix}=T^{\mathrm{f}}_{\mathrm{circ}}\begin{pmatrix}I_{+} \\ I_{-}\end{pmatrix},$$
By using Eqs.~(\ref{EQ_LinBackward}) and
(\ref{EQ_T_CIRC_F}) it becomes obvious that the T-matrix for backward propagation is given by:
\begin{equation}\hat{T}^{\mathrm{b}}_{\mathrm{circ}}=\begin{pmatrix}T_{++} & -T_{-+} \\ -T_{+-} & T_{--} \end{pmatrix}.\label{EQ_T_CIRC_B}
\end{equation}
Note that the matrix $\hat{T}^\mathrm{b}$ in an arbitrary base is not
simply given by Eq.~(\ref{EQ_LinBackward}), i.e. by interchanging
the negative off-diagonal elements but by applying the corresponding
change of basis matrix $\hat{\Lambda}$ to $\hat{T}^\mathrm{f}$ and $\hat{T}^\mathrm{b}$ in
the linear base individually.

\subsection{Asymmetric Transmission}
Although not having discussed any symmetry property at all, we want
to discuss at this point the special effect of asymmetric transmission
which attracted considerable interest due to its counter-intuitive
occurrence and discuss peculiarities related to a change of the base. The
difference of the T-matrices for opposite propagation directions is the
key to that asymmetric transmission. By asymmetric transmission $\Delta$
we understand the difference in the modulus of the total transmission
between forward and backward propagation (see Fig.~\ref{FIG_KOS}) for a certain base vector, e.g.
$\mathbf{i}=i_1\mathbf{e}_1$:
$$\Delta=|T_{11}^{\textrm{f}}|^2+|T_{12}^{\textrm{f}}|^2-|T_{11}^{\textrm{b}}|^2-|T_{12}^{\textrm{b}}|^2.$$
This quantity obviously depends on the chosen base, e.g., for a
linear state coinciding with the coordinate axis we have
($\mathbf{i}=i_x\mathbf{e}_x$):
$$\Delta^{\textrm{lin}}=|B|^2-|C|^2,$$
whereas in the circular base we have
($\mathbf{i}=i_+\mathbf{e}_+$):
$$\Delta^{\textrm{circ}}=|T_{-+}|^2-|T_{+-}|^2\neq \Delta^{\textrm{lin}}$$
in general. This dependency on the base is exploited e.g. in
[\onlinecite{FedotovPRL,NanoLettFedotov}] where asymmetric transmission for
circularly polarized light is observed without asymmetric transmission for linear
polarized light. Hence, the only proper choice is a linear
base with base vectors parallel to the principal coordinate axes. Only in
this base we can distinguish asymmetric transmission due to the structure
from asymmetric transmission due to the chosen base.

\subsection{The Eigenpolarizations}

To characterize the different structures it is useful to determine
the eigenstates of the polarization because they are uniquely
related to the symmetry. Therefore, a simple eigenvalue problem has
to be solved:
\begin{equation}
\begin{pmatrix}A & B\\ C & D \end{pmatrix}\begin{pmatrix}I_{x}\\ I_{y}\end{pmatrix}=\kappa\begin{pmatrix}I_{x}\\ I_{y}\end{pmatrix},\label{EQ_EigValProb}
\end{equation}
with the eigenvalue $\kappa$. By solving these equations we
obtain:
\begin{equation}
\kappa_{1,2}=\frac{1}{2}\left[(A+D)\pm\sqrt{(A-D)^2+4BC}\right],\label{EQ_EigValSol}
\end{equation}
where $\kappa_{1,2}$ gives the complex transmission for the
eigenstates. The eigenpolarizations are then given by simply
inserting $\kappa_{1,2}$ into Eq.~(\ref{EQ_EigValProb}) and solving for $I_x$ and $I_y$.
The eigenbasis in matrix form can be written as
\begin{equation}
\hat{\Lambda}=\begin{pmatrix}1 & 1\\ \frac{\kappa_1-A}{B} &
\frac{\kappa_2-A}{B} \end{pmatrix}\,\label{EQ_EigBaseMatrix1}
\end{equation}
with
\begin{equation}
\mathbf{i}_1=\begin{pmatrix}1\\ \frac{\kappa_1-A}{B}\end{pmatrix}\hspace{2mm}\textrm{and}\hspace{2mm}\mathbf{i}_2=\begin{pmatrix}1\\
\frac{\kappa_2-A}{B}\end{pmatrix},\label{EQ_EigVecsGeneral1}
\end{equation}
where the eigenvectors are not normalized yet. It is important
to note that the eigenbasis depends in general on the
frequency due to the dispersive behavior of the transmission. Only
for highly symmetric structures the eigenbasis is frequency
independent as will be shown later. With the use of the
characteristic polynomial of Eq.~(\ref{EQ_EigValProb}) the matrix
$\hat{\Lambda}$ can be rewritten as
\begin{equation}\hat{\Lambda}=\begin{pmatrix}1 & \frac{\kappa_2-D}{C}\\
\frac{\kappa_1-A}{B} & 1 \end{pmatrix}=\begin{pmatrix}1 & -\frac{X}{2C}\\
\frac{X}{2B} & 1 \end{pmatrix},
\label{EQ_EigBaseMatrix2}
\end{equation}
with $X=-(A-D)+\sqrt{(A-D)^2+4BC}$. Note that the matrices $\hat{\Lambda}$ in
Eqs.~(\ref{EQ_EigBaseMatrix1}) and (\ref{EQ_EigBaseMatrix2}) are different but
both are denoted simply by $\hat{\Lambda}$ not to confuse the reader with
additional indices. They are only a concatenation of eigenvectors that are
determined up to an arbitrary complex factor. The matrix $\hat{\Lambda}$
becomes unique as soon as the eigenvectors are normalized. The fractions in
Eq.~(\ref{EQ_EigBaseMatrix2}) are complex numbers, hence we can express the
eigenbasis as:
\begin{equation}
\hat{\Lambda}=\begin{pmatrix}1 & 1 \\ R_1e^{i\varphi_1} &\frac{1}{R_2}e^{-i\varphi_2} \end{pmatrix},\label{EQ_EigBaseMatrix3}
\end{equation}
with
\begin{equation}R_1e^{i\varphi_1}=\frac{X}{2B}\hspace{2mm}\textrm{and}\hspace{2mm}R_2e^{i\varphi_2}=-\frac{X}{2C}.
\label{EQ_EigVecsRelated2Matrix}
\end{equation}
The eigenvectors are obviously orthogonal only if
$$R_1=R_2\hspace{2mm}\textrm{and}\hspace{2mm}\varphi_1+\varphi_2=
(2n+1)\pi\,\textrm{with}\,n\in\mathbb{Z}.$$
This is only the case for linear, circular and a special class of elliptical
polarization. In all other cases the eigenstates are
non-orthogonal \cite{Sudha2001,Savenkov2007,Sydoruk2010}. Note that systems with orthogonal
eigenstates are sometimes termed homogeneous systems whereas systems with
non-orthogonal states are termed inhomogeneous ones \cite{Chipman1994}.

Once the eigenstates are derived, the transmission matrix can be
determined within this eigenbase by applying the transformation
~(\ref{EQ_BaseTransform}). The corresponding T-matrix is then
diagonal. Nevertheless using the T-matrix in the eigenbase is only
appropriate and convenient if the eigenstates are orthogonal and
frequency independent.

The five different classes of periodic metamaterials that can be
distinguished are closely related to their eigenstates. These five
possible sets of eigenstates are linear, circular and elliptic ones,
whereas the elliptic ones can be separated into co-rotating,
counter-rotating and general elliptic states with no fixed relation
between $\phi_1$ and $\phi_2$. Later on we will show how the symmetry
class determines the respective eigenstate.

\section{Symmetry considerations}

By the symmetry considerations in the next subsections we will show
how the symmetry properties of the structure affect the
symmetry of the T-matrix. The arising T-matrices can
be reduced to five principal forms where in general a larger number
of distinct matrices is possible by rotating the structure by an
arbitrary angle with respect to the $z$-axis. On the other hand such rotations can
be used to remove redundant information. Rotation by an angle
$\varphi$ is accomplished by applying the following matrix
operation:
\begin{equation} D_{\varphi}=\begin{pmatrix}\cos(\varphi) & \sin(\varphi) \\
-\sin(\varphi) & \cos(\varphi)\end{pmatrix}\,\rightarrow
\hat{T}_{\mathrm{new}}=D^{-1}_{\varphi}\hat{T}
D_{\varphi}\label{EQ_ArbRotation}\end{equation}
resulting in the new T-matrix $\hat{T}_{\mathrm{new}}$ of the rotated
sample. Note that the eigenvalues of the rotated
system are invariant to this operation and are uniquely related to
the principal symmetry. The actual form of the matrices and
the derived, redundant matrices will be given later in Section IV
to keep this part consistent.

In general all complex components of the T-matrix are
different if the metamaterial does not exhibit any reflection or
rotational symmetry. If such type of symmetry exists the
components of the T-matrix must reflect that. We will therefore
briefly discuss a varous symmetries and their
corresponding impact on the T-matrices.

If the metamaterial is mirror-symmetric with respect to the
$x-z-$plane, the T-matrix for the structure reflected at that
plane is identical to the original one. Therefore we have:
$$M_x=\begin{pmatrix}1 & 0 \\ 0 & -1 \end{pmatrix}:\hspace{2mm}M_x^{-1}\hat{T}^{\textrm{f}}M_x=\begin{pmatrix}A & -B \\ -C & D \end{pmatrix}=\hat{T}^{\textrm{f}}$$
\begin{equation}
\rightarrow \hat{T}^{\textrm{f}}=\begin{pmatrix}A & 0 \\ 0 & D
\end{pmatrix}\label{EQ_MX}
\end{equation}
with $M_x$ being the reflection matrix with respect to the x-axis.
So any structure that obeys that symmetry may be obviously described
by a diagonal T-matrix.

If the metamaterial is mirror-symmetric with respect to the
$y-z-$plane, we have:
$$M_y=\begin{pmatrix}-1 & 0 \\ 0 & 1 \end{pmatrix}:\hspace{2mm}M_y^{-1}\hat{T}^{\textrm{f}}M_y=\begin{pmatrix}A & -B \\ -C & D \end{pmatrix}=\hat{T}^{\textrm{f}}$$
\begin{equation}
\rightarrow \hat{T}^{\textrm{f}}=\begin{pmatrix}A & 0 \\ 0 & D
\end{pmatrix}.\label{EQ_MY}
\end{equation}
Hence, if there exists any mirror plane parallel to the $z$-axis the
T-matrix is diagonal provided that the mirror plane coincides with the
$x-$ or $y-$axes, respectively.
In such a system the eigenstates of the polarization are obviously linear states.

If the structure is $C_2$-symmetric with respect to the $z$-axis, we
have:
\begin{equation}
D_{\pi}=\begin{pmatrix}-1 & 0 \\ 0 & -1
\end{pmatrix}:\hspace{2mm}D_{\pi}^{-1}\hat{T}^{\textrm{f}}D_{\pi}=\begin{pmatrix}A
& B \\ C & D \end{pmatrix}\equiv\hat{T}^{\textrm{f}}.\label{EQ_D_PI}
\end{equation}
Hence rotating any structure by $180^\circ$ with respect to the
$z-$axis does not change the response at all.
Even if the structure does not have any further symmetry it fulfills that relation.

If the structure is $C_3$-symmetric with respect to the $z$-axis, we have:
\begin{equation}
\rightarrow \hat{T}^{\textrm{f}}=\begin{pmatrix}A & B \\ -B & A
\end{pmatrix}.\label{EQ_C3}
\end{equation}
However that symmetry is almost never met without additional  metamaterial mirror
symmetries but given here for completeness.

If the structure is $C_4$-symmetric with respect to the $z$-axis, we have:
$$D_{\frac{\pi}{2}}=\begin{pmatrix}-1 & 0 \\ 0 & -1 \end{pmatrix}:\hspace{2mm}D_{\frac{\pi}{2}}^{-1}\hat{T}^{\textrm{f}}D_{\frac{\pi}{2}}=\begin{pmatrix}D & -C \\ -B & A \end{pmatrix}=\hat{T}^{\textrm{f}}$$
\begin{equation}
\rightarrow \hat{T}^{\textrm{f}}=\begin{pmatrix}A & B \\ -B & A
\end{pmatrix}.\label{EQ_C4}
\end{equation}
Hence the structure is insensitive to linear polarized light of
any state. If there is an additional mirror-symmetry with respect to
a plane parallel or perpendicular to the $z-$axis,
the off-diagonal elements will vanish resulting in an completely
polarization independent structure. Otherwise the eigenstates will
be circularly polarized as will be shown later in detail.

Further important conclusions can be drawn by investigating the
possible mirror symmetries with respect to a plane perpendicular
to the $z-$axis. If the structure possesses this type of symmetry,
the reflected structure is the same as seen from the
backside:
$$M_x^{-1}\hat{T}^{\textrm{f}}M_x=\begin{pmatrix}A & -B \\ -C & D \end{pmatrix}=\begin{pmatrix}A & -C \\ -B & D \end{pmatrix}=\hat{T}^{\textrm{b}}$$
\begin{equation}
\rightarrow \hat{T}^{\textrm{f}}=\begin{pmatrix}A & B \\ B & D
\end{pmatrix},\label{EQ_2D}
\end{equation}
i.e., the off-diagonal elements are identical. If the system
possesses a center of inversion the matrix has also this form, because
inversion is equivalent to applying a reflection and a subsequent
rotation by $\pi$, where the latter does not change the
response as shown in Eq.~(\ref{EQ_D_PI}).

By comparison with Eq.~(\ref{EQ_C4}) it is obvious that the T-matrix
will have the form $\hat{T}^{\textrm{f}}=diag\{A,A\}$ if the structure is
additionally $C_4$-symmetric with respect to the $z-$axis.

That important relation (Eq.~(\ref{EQ_2D})) is valid for all truly
two-dimensional (planar) structures and all structures that
possess any mirror plane perpendicular to a coordinate axis,
i.e. achiral structures. In general, any substrate will break this
symmetry \cite{APLChiral,Maslovski2009,Bai2007}, but usually the substrate effect is negligible compared
to the effect of anisotropy \cite{PRBRanjan}.

Most important for our investigations are structures that cannot
be mapped onto their mirror image by proper rotations. Those structures are
called chiral. In general, the components of the T-matrix for those
structures are all different. In the context of the basic
geometry analyzed here there exist only two exceptions. The first one is
already discussed within the context of Eq.~(\ref{EQ_C4}). The
second one is an $C_2$-symmetry with respect to the $x-$ or $y-$axis.
For this type of symmetry the structure is identical from both
sides, hence
\begin{equation}
\hat{T}^{\textrm{f}}=\hat{T}^{\textrm{b}}=\begin{pmatrix}A & B \\ -B & D
\end{pmatrix}.\label{EQ_3DChiral}
\end{equation}

\section{Examples and Classification}
To understand the usefulness of the approach presented, we will
discuss the different symmetry classes for simple examples. The metaatoms
exemplarily shown in the following are assumed to be periodically arranged
in $x-$ and $y$-direction. Importantly, the symmetry constraints applied to the
unit cell have to be consistent with the symmetry of the lattice. That is
crucial since e.g. even an achiral metaatom, can result in a chiral structure
by a proper arrangement on a periodic lattice \cite{TiltedCross}.

\subsection{Simple anisotropic media}
The most significant symmetry is that of reflection symmetry with
respect to the $x-$ or $y-$axis or both. As already explained
within the context of Eqs.~(\ref{EQ_MX}) and (\ref{EQ_MY}), thes
T-matrix is then diagonal. The eigenvalues are simply $\kappa_1=A$
and $\kappa_2=D$. The eigenstates are linear states parallel and
orthogonal to the mirror plane, respectively. Only a dichroitic behavior
will be obtained and no polarization rotation occurs for light being
parallel or orthogonal to the mirror planes. If the coordinate system
is not aligned parallel to the mirror plane, the T-matrix
\begin{figure}[h]
\begin{center}
\includegraphics[width=84mm,angle=0] {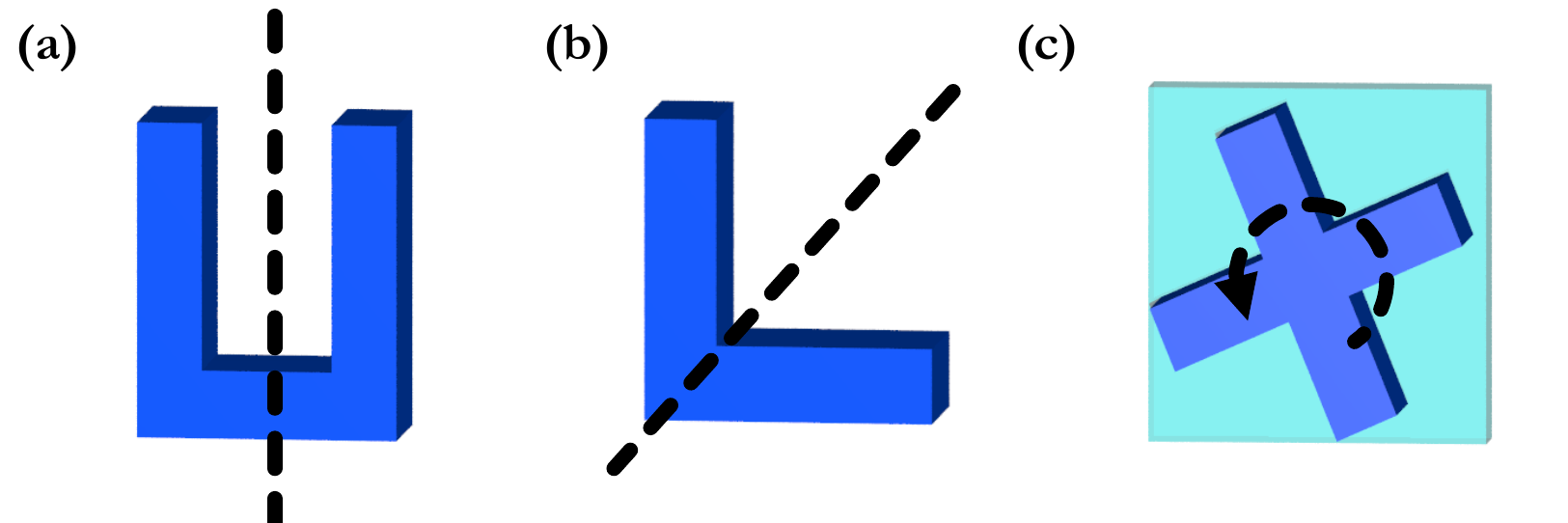}
\caption{Examples for simple anisotropic (a,b) and simple chiral (c) metaatoms.
The structures are located in the $x-y$ plane with light impinging normally to the
structure in $z$-direction. The black dashed lines indicate the mirror planes and
the rotation axis respectively. (a) Split ring resonator with mirror plane parallel
to the $y$-axis. (b) L-shaped particle with identical arms with mirror plane 45$^\circ$
inclined. (c) Cross on substrate with $C_4$ rotational symmetry with respect
to the $z$-axis. The square shaped substrate indicates the arrangement on a square
lattice, necessary for the $C_4$ of the entire system. Such an arrangement gives
rise to so-called structural chirality although the particle itself is achiral.}
\label{FIG_EXP1}
\end{center}
\end{figure}
for that system will have off-diagonal elements, which disappear after a proper
rotation. The most general form of
the T-matrix for systems with linear eigenstates is
$$\hat{T}^{\textrm{f}}=\begin{pmatrix}A & B \\ B & D \end{pmatrix},$$
but in this case the components $A,B$ and $D$ are not independent but
connected by trigonometric functions as is clear by explicitly
evaluating Eq.~(\ref{EQ_ArbRotation}) for a diagonal matrix.

An example for such a metamaterial is shown in Fig.~\ref{FIG_EXP1}~(a).
Other examples are the fishnet \cite{ShuangFishnet} and its variations \cite{FishnetVariations}, cut wire pairs \cite{NatPhotShalaev} and similar structures.
In Fig.~\ref{FIG_EXP1}~(b) we have shown a special example of a structure
with a symmetry plane which is 45$^\circ$ inclined with respect to both the $x$- and $y$-axis.
In this case the T-matrix has the form:
$$\hat{T}^{\textrm{f}}=\begin{pmatrix}A & B \\ B & A \end{pmatrix}.$$
The eigenstates are linearly orthogonal polarized, hence a rotation by
an angle $\varphi=45^\circ$ leads to a diagonal form:
$$\hat{T}^{\textrm{f}}_{\textrm{new}}=D^{-1}_{\frac{\pi}{4}}\hat{T}^{\textrm{f}}D_{\frac{\pi}{4}}=\begin{pmatrix}A' & 0 \\ 0 & D'
\end{pmatrix}=\begin{pmatrix}A+B & 0 \\ 0 & A-B \end{pmatrix}.$$
A similar structure obeying the same relations is that published in [\onlinecite{DeckerCouplingSRR}].
There, the unit cell consisting of four split-ring resonators has no rotational symmetry. But reflecting
the structure at a plane diagonal to the given unit cell leads to a structure
that is shifted by half a period in $x-$ or $y-$direction. Due to the invariance of
the optical response for periodic systems to any translation, this mirror plane leads
in fact to linearly polarized eigenstates. Therefore, rotating the structure by 45$^\circ$ results in
a diagonal T-matrix.

\subsection{Simple chiral media}
The second important group are those structures exhibiting
$C_4$-symmetry but without any additional reflection symmetry. The
T-matrix is then given by Eq.~(\ref{EQ_C4}). Since these matrices are
invariant to an arbitrary rotation $D_{\varphi}$, Eq.~(\ref{EQ_C4}) is already
the most general form of the T-matrix in a linear orthogonal base for
such systems. The T-matrix in the circular base is then diagonal:
$$T^{\textrm{f}}_{\textrm{circ}}=\begin{pmatrix}T_{++} & 0 \\ 0 & T_{--} \end{pmatrix}=\begin{pmatrix}A+iB & 0 \\ 0 & A-iB \end{pmatrix}.$$
Obviously the eigenpolarizations are circular states, since the
T-matrix is diagonal and the eigenvalues are simply $\kappa_1=A+
iB$ and $\kappa_2=A- iB$. The $y-$components of the eigenvectors
(Eq.~(\ref{EQ_EigVecsGeneral1})) are $I_{y,1,2}=\pm i$, i.e.
frequency independent. At such systems all effects related to circular dichroism
are observable whereas emphasis is put on the fact that circular dichroism
is in general accompanied by a difference in the phase advance for right-
(rcp) and left-circular polarized (lcp) light due to causality, i.e., the
real and imaginary part of the wavenumber for rcp and lcp differ in general \cite{DeckerCD}.

The difference $T_{++}-T_{--}=2iB$ is
given by the off-diagonal elements in the linear polarization representation and
specifies the optical rotation power. Systems obeying that
symmetry are prototypical optically active materials. Examples are
gammadions, swastikas (see Fig.~\ref{FIG_EXP1}(c)) or $C_4$-spirals.
Note that the influence of the substrate is important for planar
structures \cite{APLChiral,Maslovski2009} the rotation power of which is independent of the structure height since
it is a result of the substrate only.\\

\subsection{Generalized anisotropic media}
\begin{figure}[h]
\begin{center}
\includegraphics[width=84mm,angle=0] {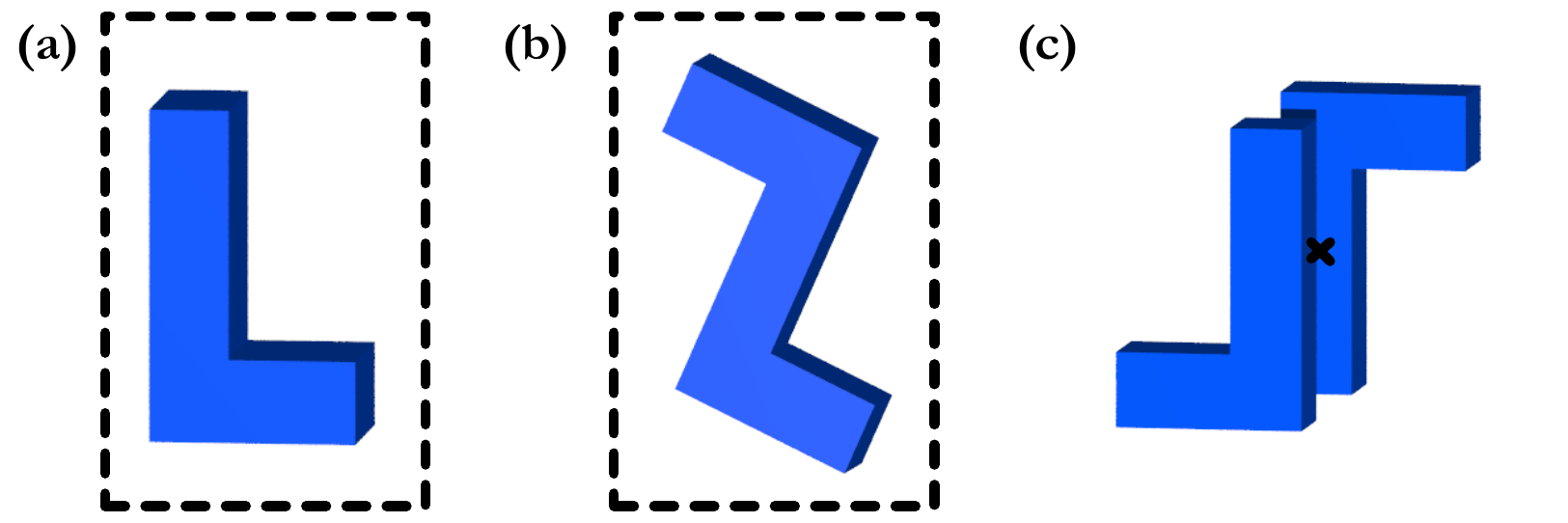}
\caption{Examples for generalized anisotropic metaatoms.
The metaatoms are located in the $x-y$ plane with light impinging normally to the
structure in $z-$direction. The black dashed lines indicate the mirror planes (a,b) and
the black cross (c) a center of inversion symmetry, respectively. (a) A planar L-shaped
metaatom with different arms. (b) A planar S-shaped metaatom with a $C_2$-symmetry with respect
to the z-axis. (c) Three-dimensional metaatom made of L-shaped particles with a center
of inversion.}
\label{FIG_AChiral}
\end{center}
\end{figure}

The third group consists of those systems that have a mirror symmetry
perpendicular to the $z-$axis or a center of inversion and at most
a $C_2$-symmetry with respect to the $z-$axis. From the latter one
we know, that it has no influence on the transmission matrix
(Eq.~(\ref{EQ_D_PI})). Examples are given in Fig.~\ref{FIG_AChiral}.
The only necessary symmetry is the reflection symmetry perpendicular
to the $z-$axis without any further restrictions. Hence, there is no
preferable alignment in the $x-y$-plane and the basic form of
the T-matrix is unaffected by any rotation with respect to the $z-$axis.

For those systems the T-matrices in the linear
and circular representation are given by:
$$T^{\textrm{f}}=\begin{pmatrix}A & B \\ B & D \end{pmatrix},\,T^{\textrm{f}}_{\textrm{circ}}=\frac{1}{2}\begin{pmatrix}A+D & A-D+2iB \\ A-D-2iB & A+D \end{pmatrix},$$
hence the eigenstates are neither linearly nor circularly states polarized.

Since we have $T_{++}=T_{--}$ there is no polarization rotation due to chirality.
In fact, it can be shown that the averaged polarization rotation accounting for
chirality vanishes in such systems \cite{DissPlum}. The off-diagonal elements in the circular
basis are different, hence the polarization conversion from left- to right-hand polarized
light and vice versa is different.
The difference in conversion is again given by the
off-diagonal elements in the linear basis $T_{+-}-T_{-+}=2iB$.
This difference is also the source of the asymmetric transmission
for circularly polarized light. Assuming (+)-polarized incident
light the total transmission $\tau$ in forward direction is
$\tau^{\textrm{f}}=|T_{++}|^2+|T_{-+}|^2$, whereas for the backward direction
we have $\tau^{\textrm{b}}=|T_{++}|^2+|T_{+-}|^2$ due to Eq.~(\ref{EQ_2D}).
Therefore, the difference in the total transmission is determined by $B$.
For (-)-polarized incident light, the results are identical. Note that there
is no asymmetric transmission for linear polarized light, as $\hat{T}^{\textrm{f}}$ is
symmetric.

It is important to note that the moduli of the off-diagonal elements are in general
in the order of those of the diagonal elements ($10^{-1}$). Hence the asymmetric transmission
can become quite large. As already indicated before, any substrate will break the mirror
symmetry in $z-$direction resulting in $B\neq C,\,|B-C|\ll |B|,\,B\approx C$.
As this difference due to the small effect of the substrate is
very weak (typically $10^{-3}$), it is often neglected and hardly measurable compared
to the asymmetric transmission effect.

The eigenstates for such a system are elliptical, co-rotating
states, as discussed e.g. in [\onlinecite{Plum2009b}]. The effects of light propagating
through such structures can be understood in terms of the concept of elliptical dichroism \cite{Zhukovsky2009}.
By using Eqs.~(\ref{EQ_EigBaseMatrix3}), \ref{EQ_EigVecsRelated2Matrix} and $B=C$
they can be expressed in normalized form as
$$\mathbf{i}_1=\frac{1}{\sqrt{1+R^2}}\begin{pmatrix}1 \\ Re^{i\varphi}\end{pmatrix},\,\mathbf{i}_2=\frac{R}{\sqrt{1+R^2}}\begin{pmatrix}1 \\ -\frac{1}{R}e^{-i\varphi}\end{pmatrix}.$$
They are only orthogonal for $\varphi=n\pi$ with $n\in\mathbb{N}$
leading to linear eigenstates.

Note that planar structures with that symmetry can be described by
an effective permittivity tensor independent on the wavevector, i.e.
without magnetoelectric coupling \cite{Petschulat2010}. That is why we call this
group generalized anisotropic structures.

The most general form is again obtained by applying a rotation by an
arbitrary angle $\varphi$ leading to
$$\hat{T}^{\textrm{f}}_{\textrm{new}}=D^{-1}_{\varphi}\hat{T}^{\textrm{f}}D_{\varphi}=\begin{pmatrix}A' & B' \\ B' & D' \end{pmatrix},$$
hence the general form is invariant since no preferred alignment exists.

\subsection{Generalized chiral media}

The forth group are chiral structures that have an additional
$C_2$-symmetry with respect to the $x-$ or $y-$axis. The T-matrix
obeys the form:
\begin{equation}
T^{\textrm{f}}=\begin{pmatrix}A & B \\ -B & D \end{pmatrix},\,T^{\textrm{f}}_{\textrm{circ}}=\frac{1}{2}\begin{pmatrix}A+D+2iB & A-D \\ A-D & A+D-2iB \end{pmatrix},
\label{EQ_TMatrixChiral}
\end{equation}
hence there is no difference in the polarization conversion and
hence no asymmetric transmission neither for linear nor for
circular polarized light. Furthermore there is obviously no
asymmetric transmission in any base, since the structure is
identical from both sides when the axis of rotation coincides
with the $x-$ or $y-$axis.

But there is a difference in the quantity $T_{++}-T_{--}=2iB$ determining
\begin{figure}[h]
\begin{center}
\includegraphics[width=84mm,angle=0] {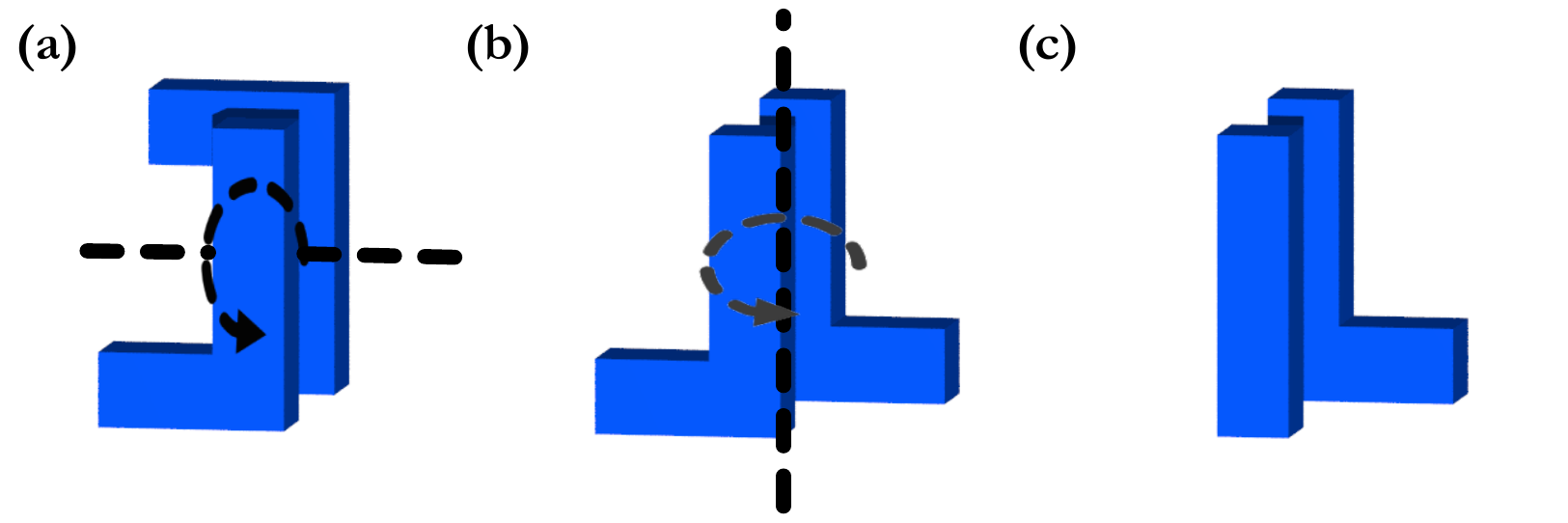}
\caption{Examples for generalized chiral metaatoms (a,b) and a no-symmetry metaatom (c).
The structures are located in the $x-y$ plane with light impinging normally to the
structure in $z-$direction. The black dashed lines indicate the axes of rotational symmetry (a,b).
(a,b) Three-dimensional structures made of two L-shaped particles with $C_2$-symmetry with respect to
the $x-$ or $y-$axis, respectively. They are identical for forward and backward propagation.
(c) A three-dimensional structure made of an L-shaped particle and an I-shaped particle with
no symmetry at all.}
\label{FIG_CHIRAL3D}
\end{center}
\end{figure}
the optical rotation power typical for chiral structures. In
contrast to the second group we have an additional anisotropy
($A\neq D$) hence the eigenstates are not circular but
elliptical counter-rotating. Again, by using
Eqs.~(\ref{EQ_EigBaseMatrix3}), (\ref{EQ_EigVecsRelated2Matrix}) and
$C=-B$ they can be expressed in normalized form as
$$\mathbf{i}_1=\frac{1}{\sqrt{1+R^2}}\begin{pmatrix}1 \\ Re^{i\varphi}\end{pmatrix},\,\mathbf{i}_2=\frac{R}{\sqrt{1+R^2}}\begin{pmatrix}1 \\ \frac{1}{R}e^{-i\varphi}\end{pmatrix}.$$
They are only orthogonal if $\varphi=\frac{\pi}{2}+n\pi$ with $n\in\mathbb{N}$
leading to circular counter-propagating eigenstates typical for chiral structures.
That is why we term this group generalized chiral structures.

Typical examples are shown in Fig.~\ref{FIG_CHIRAL3D} a) and b).
Another important example are three-dimensional spirals
\cite{Lee2005,Gansel2010,Silveirinha2008,Hodgkinson2000,Lakthakia2000} with $\frac{N}{2}$
whorls aligned along the $z-$axis. Spirals with integer whorls are clearly identical
for both propagation directions, whereas spirals with half-integer whorls
are identical after rotation by $\pi$ around the $z-$axis keeping the
response unaffected.

Note that for an arbitrary rotation $D_{\varphi}$ all matrix elements
are different, hence the symmetry axis must be aligned with a principal
coordinate axis to achieve the form of Eq.~(\ref{EQ_TMatrixChiral}).
In particular if the system is rotated by $45^\circ$ the T-matrix has the form:
$$T^{\textrm{f}}=\begin{pmatrix}A' & B' \\ C' & A' \end{pmatrix}.$$
Nevertheless if the eigenvectors of the arbitrarily oriented system are
elliptical counter-rotating the convenient form of Eq.~(\ref{EQ_TMatrixChiral})
can be achieved by a proper alignment of the system.

\subsection{Arbitrary complex media}

The fifth and last group are chiral structures without any symmetry. A simple example is shown in Fig.~\ref{FIG_CHIRAL3D} c). Here
all elements of the T-matrices in the linear as well as in the circular base are
different:
\begin{equation}
T^{\textrm{f}}=\begin{pmatrix}A & B \\ C & D \end{pmatrix}.
\label{EQ_TMatrixArbitraryComplex}
\end{equation}
It is impossible to achieve $|B|=|C|$ by a proper rotation. Therefore,
independent of the base asymmetric transmission occurs always and in
particular also for linearly polarized light. All effects of generalized
anisotropy as well as generalized chirality can be observed. The normalized
eigenvectors can be expressed as:
$$\mathbf{i}_1=\frac{1}{\sqrt{1+R_1^2}}\begin{pmatrix}1 \\ R_1e^{i\varphi_1}\end{pmatrix},\,\mathbf{i}_2=\frac{R_2}{\sqrt{1+R_2^2}}\begin{pmatrix}1 \\ \frac{1}{R_2}e^{-i\varphi_2}\end{pmatrix}.$$
whereas $R_1(\omega)\neq R_2(\omega)$ and $\varphi_1(\omega)\neq\varphi_2(\omega)$.
The eigenstates are strongly depending on the actual value of the components
of $\hat{T}$ and are simply elliptical, whereas no principal rotation direction
is assignable. Linear as well as elliptical counter- and co-rotating states and
combinations of them with no fixed phase relation can be found in general.

An example of such an structure is investigated in detail both numerically and
experimentally in [\onlinecite{PRLMenzel}].

\section{Summary}
A summarizing overview of possible structures and the corresponding basic
forms of the T-matrices are shown in Table \ref{TABLE_ORDER}. Once the general
form of the T-matrix is known all effects regarding the observable polarization
phenomena can be fully deduced. Based on our investigations it is easy to provide an algorithm to
determine the general form of the T-matrix for an unknown sample by measuring
transmitted intensities with the help of linear polarizers only.

A possible
approach can be as follows:
\\
\begin{table}[h]
\extrarowheight3pt
\begin{tabular}{||p{18mm}|c|c|p{20mm}||}
  \hline\hline
  \textbf{symmetry} & \textbf{examples} & \textbf{T-matrix} & \textbf{eigenstates} \\ \hline\hline
  $M_{xz}$ ($M_{yz}$)& \includegraphicstotab[width=15mm]{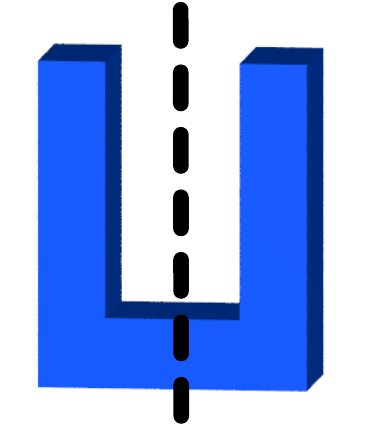} & $T=\begin{pmatrix}A & 0 \\ 0 & D \end{pmatrix}$ & linear \\ \hline
  $C_{4,z}$ ($C_{3,z}$) & \includegraphicstotab[width=15mm]{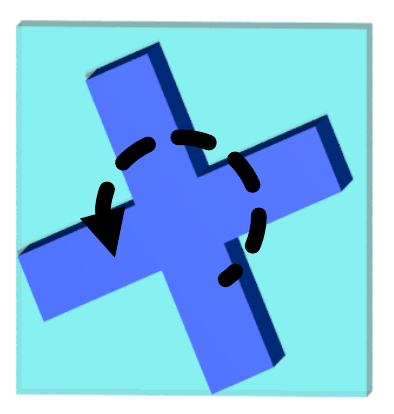} & $T=\begin{pmatrix}A & B \\ -B & A \end{pmatrix}$ & circular \\ \hline
  $M_{xy}$ ($C_{2,z}$, inversion symmetry)& \includegraphicstotab[width=15mm]{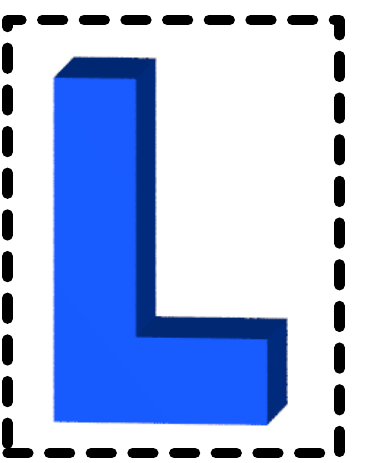} & $T=\begin{pmatrix}A & B \\ B & D \end{pmatrix}$ & elliptic,\newline co-rotating \\ \hline
  $C_{2,y}$ ($C_{2,x}$) & \includegraphicstotab[width=15mm]{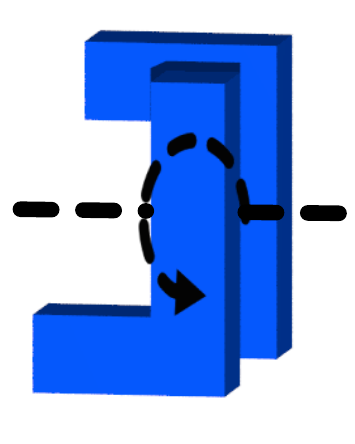} & $T=\begin{pmatrix}A & B \\ -B & D \end{pmatrix}$ & elliptic, counter-rotating \\ \hline
  no symmetry ($C_{2,z}$)& \includegraphicstotab[width=15mm]{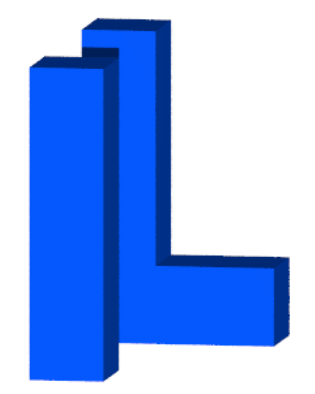} & $T=\begin{pmatrix}A & B \\ B & D \end{pmatrix}$ & elliptic \\
  \hline\hline
\end{tabular}
\caption{Overview of possible symmetries, typical metaatoms, the corresponding T-matrices
and their eigenstates of the polarization. For every symmetry group only a single example is
shown. Other possible symmetries resulting in the same type of T-matrices are given in
brackets. Here $M_{ij}$ designates mirror symmetry with respect to the  $ij-$plane and $C_n,i$
means n-fold rotational symmetry with respect to the $i-$axis.}%
\label{TABLE_ORDER}%
\end{table}
1. Use linearly polarized light and measure the orthogonally polarized output
while rotating the sample. If the output vanishes for every rotation angle the
medium is polarization independent, i.e., a simple isotropic medium. If the
output vanishes for some rotation angles and this angle is independent of the wavelength,
the structure is simple anisotropic. If no such rotation angle can be found
there is obviously no mirror plane parallel to the $z$-axis.\\
2. If the transmitted intensity is independent of the rotation of the sample for both co- and
cross-polarized light, the eigenstates are circular polarized and the structure is simple chiral.\\
3. If both aforementioned procedures do not provide a positive results, the structure
is more complex and the measurements become more difficult too. To distinguish between
the remaining possible forms it is necessary to measure the off-diagonal entries
of the T-matrix simultaneously. If these off-diagonal elements are identical for a
fixed wavelength and a fixed rotation angle independent of their particular choice,
the structure is generalized anisotropic. If the off-diagonal elements are identical
only for a fixed rotation angle but for every wavelengths the structure is generalized
chiral. In all other cases we have $A\neq B \neq C \neq D$.\\
By using circular polarized light a similar scheme can be obtained however it would require circular analyzers as well.

\section{Conclusion}
Taking advantage of symmetry considerations we have analyzed the potential of various MMs to affect the polarization state of light upon transmission. By focusing the attention on any optical response that is directly accessible in an experiment, the properties of MMs may become so involved that the establishment of valid constitutive relation may be beyond what is possible for structures with an ever increasing complexity. We have explicitly shown that all MMs belong to one of five different classes; each being characterized by certain relations that connect the entries of the T-matrix and each class is able to support specific polarization phenomena. The sub-wavelength nature of MMs is the only requirement for these considerations. Moreover, the symmetry operations applied to the metaatoms have to be consistent with the symmetry of the lattice and it is required that the MM is sandwiched between identical media. Nonetheless, we have explicitly listed all relevant structures where a violation of this assumption causes deviations.  To foster practical application of this classification we have finally provided a protocol useful to reveal the underlying symmetry of an unknown MM and its T-matrix from far-field measurements of the transmitted intensities only. Once it is identified, all the achievable optical properties that affect the state of polarization are fully disclosed.

\section*{Acknowledgements}
We acknowledge financial support from the
German Federal Ministry of Education and Research (Metamat and PhoNa) and by
the Thuringian State Government (MeMa).

\end{document}